\documentclass[aps, prd, superscriptaddress, nofootinbib, preprintnumbers, twocolumn, showpacs]{revtex4}
\usepackage{amsmath}

\begin{document}

\title{%
  \hfill{\normalsize\vbox{%
  }}\\
  \vspace{-0.5cm}
  {\bf Implications of Holographic QCD \\
  in ChPT with Hidden Local Symmetry}
 } 
\author{{\bf Masayasu Harada}}
\author{{\bf Shinya Matsuzaki}}
\author{{\bf Koichi Yamawaki}}

\affiliation{ Department of Physics, Nagoya University,
Nagoya, 464-8602, Japan,}

\begin{abstract}
Based on the chiral perturbation theory (ChPT) 
with the hidden local symmetry, 
we propose a methodology to calculate a part of the large $N_c$ corrections
in the holographic QCD (HQCD). 
As an example, 
we apply the method to an HQCD model recently proposed 
by Sakai and Sugimoto. 
We show that the $\rho$-$\pi$-$\pi$ coupling 
becomes in good agreement with the experiment due to 
the $1/N_c$-subleading corrections. 
\end{abstract}

\maketitle

\section{Introduction}

Recently the duality in string/gauge theory \cite{Maldacena:1997re} has provided us with 
a new perspective for solving the problem of strongly coupled gauge theories:
Strongly coupled gauge theory can be reformulated from the weakly coupled string theory based on 
the AdS/CFT correspondence~\cite{AdS/CFT}. 
Some important qualitative features of the dynamics of QCD such as the confinement and 
chiral symmetry breaking have been reproduced from this holographic point of view, 
so-called holographic QCD (HQCD), although the theory in the UV region is substantially different from QCD, 
i.e., lack of asymptotic freedom. 
Several authors~\cite{SaSu,HQCD} proposed a model of HQCD where the chiral symmetry breaking is realized. 
In particular, starting with a stringy setting, Sakai and Sugimoto (SS)~\cite{SaSu} have succeeded in producing the realistic chiral symmetry breaking 
$U(N_f)_L\times U(N_f)_R$ down to $U(N_f)_V$ and also a  natural emergence of 
the hidden local symmetry (HLS)~\cite{Bando:1984ej} for vector/axialvector mesons. 
Moreover, most of them~\cite{SaSu,HQCD} analyze observables of QCD related to 
the pion and the vector mesons in the large $N_c$ limit such as 
$m_{\rho}^2/(g_{\rho\pi\pi}^2F_{\pi}^2) \simeq 3.0$, where $g_{\rho\pi\pi}$ denotes the $\rho$-$\pi$-$\pi$ coupling, 
and $F_{\pi}$ does the pion decay constant. 
This, however, 
substantially deviates
from one of the celebrated KSRF relations (KSRF II), 
$m_{\rho}^2/(g_{\rho\pi\pi}^2F_{\pi}^2) = 2$, which agrees with the experiment. 
Since the holographic result is only the one at the leading order in the $1/N_c$ expansion, 
the deviation may be cured by subleading effects in the $1/N_c$ expansion. 
So far, however, no holographic models succeed in including the effect of the $1/N_c$ corrections. 
On the other hand, it is well known that  meson loops yield the next order in  the $1/N_c$ expansion.

In this paper
\footnote{
          The result of this paper was presented at the annual meeting
          of the Physical Society of Japan (held at Tokyo University of Science at Noda), 
          March 24-27, 2005, and at a workshop on ``Progress in the Particle Physics 2005'' held at 
          Yukawa Institute for Theoretical Physics, Kyoto University, June 20-24, 2005. 
          }  
we propose a methodology for calculating a part of the $1/N_c$ corrections to the HQCD through 
the meson loop, based on the ``HLS chiral perturbation theory (ChPT)'' 
\cite{Harada:1992bu,Tanabashi:1993sr,Harada:1999zj,HY:PRep} which incorporates 
the vector meson loops into the ChPT~\cite{Gas:84} through the HLS model \cite{Bando:1984ej}. 
It is important to note \cite{Georgi:1989gp} that the {\it HLS is crucial for the systematic power counting} 
when the vector meson mass is light (see, for a review, Ref.\cite{HY:PRep}). 
As an example, 
we apply our method to an HQCD model proposed by SS~\cite{SaSu}. 
We show that the $1/N_c$ corrections make
the ratio $m_{\rho}^2/(g_{\rho\pi\pi}^2F_{\pi}^2)$
in good agreement with the experimental value or the KSRF II relation.
Our formalism proposed in this paper is applicable to other models holographically dual to 
strongly coupled gauge theories, which will give us implications of HQCD.

\section{Review of a holographic model}

Let us start with the low-energy effective action on the 5-dimensional space-time 
induced from a holographic model, based on the $N_f$ D8-$\overline{\rm D8}$ branes 
transverse to the $N_c$ D4-branes, proposed by the authors in Ref.~\cite{SaSu}: 
\begin{eqnarray}
S_{D8}&=& N_c G  
        \int d^4x dz  
         \Bigg( 
               - \frac{1}{2} K^{-1/3}(z) {\rm tr}[ 
                                                   F_{\mu\nu} F^{\mu\nu} 
                                                 ] 
\nonumber \\ 
         &&     
               +      K(z) M_{\rm KK}^2  {\rm tr }[ 
                                                   F_{\mu z} F^{\mu z} 
                                                  ] 
               + \mathcal{O}(F^3) 
         \Bigg)
\,, \label{D8 brane action}
\end{eqnarray} 
where $K(z)$ is the induced measure of 5-dimensional space-time given by 
\begin{equation} 
K(z) = 1 + z^2 
\,. 
\end{equation} 
The coupling $G$ is the rescaled 't~Hooft coupling expressed as 
\begin{equation} 
G = \frac{N_cg_{\rm YM}^2}{108 \pi^3} 
\,, 
\end{equation}
where $g_{\rm YM}$ is the gauge coupling of 
the $U(N_c)$ gauge symmetry on the $N_c$ D4-branes. 
It should be noted that the mass scale $M_{KK}$ in Eq.(\ref{D8 brane action}) is related to 
the scale of the compactification of the $N_c$ D4-branes onto the $S^1$.

The five dimensional gauge field $A_{M}$ transforms as 
\begin{eqnarray}
A_{M}(x^{\mu},z) 
  & \rightarrow & 
  g(x^{\mu},z) \cdot A_{M}(x^{\mu},z) \cdot g^{\dagger}(x^{\mu},z) 
  \nonumber \\ 
  && 
   - i \partial_{M} g(x^{\mu},z) \cdot g^{\dagger}(x^{\mu},z) 
\,, 
\end{eqnarray} 
where $g(x^{\mu}, z)$ is the transformation matrix of the 5-dimensional gauge symmetry. 
We choose the same boundary condition of the 5-dimensional gauge field $A_{M}$ as done in 
Ref.~\cite{SaSu}: 
\begin{equation} 
A_{M}(x^{\mu},z = \pm \infty) = 0 
\,, 
\end{equation}
which makes the local chiral symmetry be 
a global one $g_{R,L}\in \mbox{U}_{R,L}(N_f)$. 
The chiral field $U$ defined in Ref.~\cite{SaSu}: 
\begin{equation}
U(x^{\mu})
   = 
   {\rm P} \exp \left[ i \int_{-\infty}^{\infty} dz' A_z(x^{\mu},z') \right] 
\, 
\end{equation}
is parameterized by the NG boson field $\pi$ as 
\begin{equation} 
 U(x^{\mu}) = e^{\frac{2i \pi(x^{\mu})}{F_{\pi}}} 
\, , 
\end{equation} 
where $F_\pi$ denotes the decay constant of $\pi$. 
$U$ is divided as 
\begin{equation} 
 U(x^{\mu}) = \xi_L^{\dagger}(x^{\mu}) \cdot \xi_R(x^{\mu}) 
\,, 
\end{equation} 
where 
\begin{eqnarray} 
\xi_{R,L}(x^{\mu}) 
   = 
   {\rm P} \exp \left[ i \int_{0}^{\pm\infty} dz' A_z(x^{\mu},z')  \right] 
\,.
\end{eqnarray}
$\xi_{R,L}$ transform as~\cite{Bando:1984ej} 
\begin{equation} 
\xi_{R,L}^\prime = h(x^{\mu}) \cdot \xi_{R,L} \cdot g_{R,L}^{\dagger} 
\, , 
\end{equation} 
where $h(x^\mu)=g(x^{\mu}, z=0)$ is a transformation of the HLS. 
We further parameterize $\xi_{R,L}$ as~\cite{Bando:1984ej} 
\begin{equation} 
\xi_{R,L}(x^\mu) = e^{\frac{i\sigma(x^\mu)}{F_\sigma}} \cdot e^{\pm \frac{i\pi(x^\mu)}{F_\pi}} 
\,, 
\end{equation} 
 where $\sigma$ denote the NG bosons associated with the spontaneous breaking of the HLS, and 
$F_\sigma$ the decay constant of $\sigma$. 
The $\sigma$ are absorbed into the gauge bosons of the HLS which acquire the mass 
through the Higgs mechanism.

It is convenient to work in the $A_z(x^\mu,z)\equiv 0$ gauge~\cite{SaSu}. 
There still exists a residual gauge symmetry, 
$h(x^\mu)=g(x^{\mu}, z=0)$, 
which was identified with the hidden local symmetry (HLS) in Ref.~\cite{SaSu}. 
The 5-dimensional gauge field $A_{\mu}$ transforms 
under the residual gauge symmetry (HLS) as 
\begin{eqnarray}
A_{\mu}(x^{\mu},z) 
 &\rightarrow& 
       h(x^{\mu}) \cdot A_{\mu}(x^{\mu},z) \cdot h^{\dagger}(x^{\mu})  
 \nonumber \\      
  &&  \hspace{20pt}     
       - i \partial_{\mu} h(x^{\mu}) \cdot h^{\dagger}(x^{\mu}) 
\,. \label{trans amu}
\end{eqnarray} 
In this gauge the NG boson fields are included in the boundary condition for 
the 5-dimensional gauge field $A_{\mu}$ as 
\begin{equation} 
A_{\mu}(x^{\mu}, z= \pm \infty) = \alpha^{R,L}_{\mu}(x^{\mu}) 
\,, 
\end{equation} 
where 
\begin{equation} 
\alpha^{R,L}_{\mu} (x^{\mu}) = i \xi_{R,L} (x^{\mu}) \partial_{\mu} \xi^{\dagger}_{R,L}(x^{\mu})
\, , 
\end{equation} 
which transform under the HLS as in the same way as in Eq.~(\ref{trans amu}).

\section{Relation to HLS in large $N_C$ limit}

In contrast to Ref.~\cite{SaSu} where vector meson fields are identified with 
the CCWZ matter fields transforming {\it homogeneously} under HLS, 
we here introduce vector meson fields as an infinite tower of the HLS {\it gauge fields} 
$V_\mu^{(k)}$ ($k=1,2,\ldots$), which transform {\it inhomogeneously} under the HLS 
as in Eq.~(\ref{trans amu})~\cite{HY:PRep}. 
Using $V_\mu^{(k)}$ together with $\alpha^{R,L}_{\mu}$,
we expand the 5-dimensional gauge field $A_{\mu}$ as 
\begin{eqnarray}
A_{\mu}(x^{\mu},z) 
&=& 
\alpha^R_{\mu}(x^{\mu}) \phi^r(z) + 
\alpha^L_{\mu}(x^{\mu}) \phi^l(z) 
\nonumber \\ 
&& 
+ 
\sum_{k \ge 1} V^{(k)}_{\mu}(x^{\mu}) \phi_k(z) 
\, , \label{A expansion}
\end{eqnarray}
where the functions $\phi^r$, $\phi^l$ and $\phi_k$ ($k=1,2,\ldots$) form a complete set 
in the $z$-coordinate space. 
These functions \{$\phi^r$, $\phi^l$, $\phi_k$\} are different from the eigenfunctions $\psi_n$ in \cite{SaSu} 
which satisfy the eigenvalue equation 
\begin{equation} 
 -K^{1/3} \partial_z  \left( K \partial_z \psi_n \right) = \lambda_n \psi_n  
\, ,
\end{equation}  
with the eigenvalues $\lambda_n$. 
Then, the functions \{$\phi^r$, $\phi^l$, $\phi_k$ \} are not separately the solutions of the eigenvalue equation 
but are expressed by linear combinations of the solutions, as we will see later.

Substituting Eq.(\ref{A expansion}) into the action (\ref{D8 brane action}), 
we obtain the 4-dimensional theory with an infinite tower of the massive vector and axialvector mesons and 
the NG bosons associated with the chiral symmetry breaking. 
We would like to stress that, since the 5-dimensional gauge field $A_\mu$ is expanded in terms of 
the HLS gauge fields $V_{\mu}^{(k)}$, the action (\ref{D8 brane action}) is expressed as the form 
{\it manifestly gauge invariant} under the HLS, which enables us to calculate the $1/N_c$-subleading correction 
in a systematic way.

Let us concentrate on the lightest vector meson together with the NG bosons by integrating out 
the heavy vector and axialvector meson fields
\footnote{ 
            This is contrasted with simply putting the heavy fields  
            $V_{\mu}^{(k)}$ $(k \ge 2) =0 $ in Eq.(\ref{A expansion}). 
            The wave functions $\phi_k(z)$ are thus modified, when we integrate out 
            the heavier fields~\cite{forthcoming}. 
            }. 
As a result, the HLS gauge field $V_{\mu}$ corresponding to the lightest vector meson is embedded into $A_{\mu}$ as 
\begin{eqnarray} 
A_{\mu}(x^{\mu},z) 
     &=&
     \alpha_{\mu}^R (x^{\mu}) \varphi^r(z) + \alpha^L_{\mu}(x^{\mu}) \varphi^l(z)  
\nonumber \\ 
     && 
     + V_{\mu}(x^{\mu}) \varphi(z)
\, , \label{A expansion1}
\end{eqnarray} 
where $\varphi^r$, $\varphi^l$ and $\varphi$ denote the wave functions modified by 
integrating out the heavier mesons. 
Note that they satisfy the following constraint: 
\begin{eqnarray}
\varphi^r(z) + \varphi^l(z) + \varphi(z) = 1 
\,, 
\label{constraint for n equal 1} 
\end{eqnarray} 
which follows from the consistency condition between the transformation properties ({\it inhomogeneous term}) of 
the left and right hand sides of Eq.(\ref{A expansion1}). 
The relations between $\{\varphi^r , \varphi^l , \varphi \}$ and 
the eigenfunctions of the eigenvalue equation are obtained in the following way: 
We introduce the 1-forms $\widehat{\alpha}_{\mu||}$ and $\widehat{\alpha}_{\mu\perp}$ defined as 
\begin{eqnarray}
{\widehat{\alpha}}_{\mu||}(x^{\mu})
      &=&
      \frac{\alpha^R_{\mu}(x^{\mu})+\alpha^L_{\mu}(x^{\mu})}{2} - V_{\mu}(x^{\mu}) 
\,,  \\
{\widehat{\alpha}}_{\mu\perp}(x^{\mu})
      &=&
      \frac{\alpha^R_{\mu}(x^{\mu})-\alpha^L_{\mu}(x^{\mu})}{2}
\,. 
\end{eqnarray} 
Then Eq.(\ref{A expansion1}) is rewritten into the following form: 
\begin{eqnarray} 
A_{\mu}(x^{\mu},z) 
      &=&
      {\widehat{\alpha}}_{\mu\perp}(x^{\mu}) \left(\varphi^r(z)- \varphi^l(z) \right) 
\nonumber \\ 
      && 
      + \left( \widehat{\alpha}_{\mu||}(x^{\mu})+V_{\mu}(x^{\mu})\right) 
      \left(\varphi^r(z)+ \varphi^l(z) \right) 
\nonumber \\
      && 
      +V_{\mu}(x^{\mu}) \varphi(z) 
\,. \label{A expansion2}
\end{eqnarray} 
Since the 1-form $\widehat{\alpha}_{\mu\perp}$ includes the NG boson field as 
$\widehat{\alpha}_{\mu \perp} = \frac{1}{F_{\pi}} \partial_{\mu} \pi + \cdots$, 
we identify the combination $\varphi^r-\varphi^l$ with the eigenfunction $\psi_0$ for 
the zero eigenvalue as 
\begin{equation} 
 \varphi^r(z) - \varphi^l(z) = \psi_0(z)= \frac{2}{\pi} \tan^{-1}z 
\,.  
\end{equation} 
On the other hand, since the HLS gauge field $V_{\mu}$ corresponds to the lightest vector meson, 
we identify the wave function $\varphi$ with the eigenfunction of 
the first excited KK mode, 
\begin{equation} 
\varphi(z) = - \psi_1(z)
\, .
\end{equation}  
Then, by using Eq.(\ref{constraint for n equal 1}), 
the wave functions $\varphi^r$ and $\varphi^l$ are expressed in terms of 
the eigenfunctions $\psi_0$ and $\psi_1$ as 
\begin{eqnarray}
\varphi^{r,l}(z)
      &=& 
      \frac{1}{2} \pm \frac{1}{2} \psi_0(z) + \frac{1}{2}\psi_1(z)
\,. 
\end{eqnarray} 
By using this, Eq.(\ref{A expansion2}) is rewritten into the following form: 
\begin{eqnarray}
A_{\mu}(x^{\mu},z) 
      &=& 
      \widehat{\alpha}_{\mu\perp}(x^{\mu}) \psi_0(z) 
      + \left( \widehat{\alpha}_{\mu||}(x^{\mu}) + V_{\mu}(x^{\mu}) \right) 
\nonumber \\ 
      && 
      + \widehat{\alpha}_{\mu||}(x^{\mu}) \psi_1(z) 
\,. 
\end{eqnarray} 
It should be noticed that neither the wave function $\varphi^r$ nor $\varphi^l$ is the eigenfunction
for the zero eigenvalue. 
This is the reflection of the well-known fact that the massless photon field is given by 
a linear combination of the HLS gauge field and the gauge field corresponding to 
the chiral symmetry~\cite{Bando:1984ej,HY:PRep}.

Now, since we introduce the vector meson field as the gauge field of the HLS, 
the derivative expansion of the Lagrangian becomes possible. 
This is an important difference compared with the formulation done in Ref.~\cite{SaSu}. 
Then, the leading Lagrangian counted as $\mathcal{O}(p^2)$ in the derivative expansion is constructed by 
the terms generated from the $F_{\mu z}F^{\mu z}$ term in the action (\ref{D8 brane action}) together with 
the kinetic term of the HLS gauge field $V_{\mu}$ from the $F_{\mu\nu}F^{\mu\nu}$ term. 
On the other hand, the $\mathcal{O}(p^4)$ terms come from the remainder of the $F_{\mu\nu}F^{\mu\nu}$ term 
in the action (\ref{D8 brane action}). 
The resultant Lagrangian takes the form of the HLS model~\cite{Bando:1984ej,HY:PRep}: 
\begin{eqnarray}
\mathcal{L}
      &=&
      F_{\pi}^2 {\rm tr}[ \widehat{\alpha}_{\mu\perp} \widehat{\alpha}^{\mu}_{\perp} ] 
      + F_\sigma^2 {\rm tr}[ \widehat{\alpha}_{\mu||} \widehat{\alpha}^{\mu}_{||} ] 
\nonumber \\ 
      && 
      - \frac{1}{2g^2}{\rm tr}[ V_{\mu\nu} V^{\mu\nu} ] 
      + \mathcal{L}_{(4)} 
\,, \label{HLS Lag} 
\end{eqnarray}
where $\mathcal{L}_{(4)}$ is constructed by the ${\mathcal O}(p^4)$ terms~\cite{Tanabashi:1993sr,HY:PRep}: 
\begin{eqnarray}
\mathcal{L}_{(4)} 
&=&
y_1 \, 
{\rm tr}[ 
           \widehat{\alpha}_{\mu\perp} \widehat{\alpha}^{\mu}_{\perp} 
           \widehat{\alpha}_{\nu\perp} \widehat{\alpha}^{\nu}_{\perp} 
        ] 
+
y_2 \, 
{\rm tr}[
           \widehat{\alpha}_{\mu\perp} \widehat{\alpha}_{\nu\perp} 
           \widehat{\alpha}^{\mu}_{\perp} \widehat{\alpha}^{\nu}_{\perp} 
         ] 
\nonumber \\
&&
+ 
y_3 \, 
{\rm tr}[ 
           \widehat{\alpha}_{\mu||} \widehat{\alpha}^{\mu}_{||} 
           \widehat{\alpha}_{\nu||} \widehat{\alpha}^{\nu}_{||} 
         ] 
+ 
y_4 \, 
{\rm tr}[ 
           \widehat{\alpha}_{\mu||} \widehat{\alpha}_{\nu||} 
           \widehat{\alpha}^{\mu}_{||} \widehat{\alpha}^{\nu}_{||} 
         ] 
\nonumber \\
&&
+ 
y_5 \, 
{\rm tr}[
           \widehat{\alpha}_{\mu\perp} \widehat{\alpha}^{\mu}_{\perp} 
           \widehat{\alpha}_{\nu||} \widehat{\alpha}^{\nu}_{||} 
         ] 
+ 
y_6 \, 
{\rm tr}[ 
           \widehat{\alpha}_{\mu\perp} \widehat{\alpha}_{\nu\perp} 
           \widehat{\alpha}^{\mu}_{||} \widehat{\alpha}^{\nu}_{||} 
         ] 
\nonumber \\ 
&& 
+ 
y_7 \, 
{\rm tr}[ 
           \widehat{\alpha}_{\mu\perp} \widehat{\alpha}_{\nu\perp} 
           \widehat{\alpha}^{\nu}_{||} \widehat{\alpha}^{\mu}_{||} 
         ] 
\nonumber \\
&& 
+ 
y_8 \, 
\Bigg\{ 
        {\rm tr}[ 
                  \widehat{\alpha}_{\mu\perp} \widehat{\alpha}^{\mu}_{||} 
                  \widehat{\alpha}_{\nu\perp} \widehat{\alpha}^{\nu}_{||} 
                 ] 
+ 
        {\rm tr}[ 
                  \widehat{\alpha}_{\mu\perp} \widehat{\alpha}^{\nu}_{||} 
                  \widehat{\alpha}_{\nu\perp} \widehat{\alpha}^{\mu}_{||} 
                 ] 
\Bigg\} 
\nonumber \\ 
&& 
+ 
y_9 \, 
{\rm tr}[ 
           \widehat{\alpha}_{\mu\perp} \widehat{\alpha}^{\nu}_{||} 
           \widehat{\alpha}_{\mu\perp} \widehat{\alpha}^{\nu}_{||} 
         ] 
\nonumber \\
&&
+ i z_4 \, 
{\rm tr}[ 
          V_{\mu\nu} \widehat{\alpha}^{\mu}_{\perp} \widehat{\alpha}^{\nu}_{\perp} 
         ] 
+ 
i z_5 \, 
{\rm tr}[ 
          V_{\mu\nu} \widehat{\alpha}^{\mu}_{||} \widehat{\alpha}^{\nu}_{||} 
         ] 
\,. 
\end{eqnarray} 
Note that all the parameters in the Lagrangian are expressed in terms of 
the parameters of the 5-dimensional gauge theory as 
\begin{eqnarray}
F_{\pi}^2 
      &=& 
      N_c GM_{KK}^2  \int dz K(z) \left[ \dot{\psi}_0(z) \right]^2
\,, \label{Fpi} \\
F_{\sigma}^2
      &=& 
      N_c GM_{KK}^2  \lambda_1 \langle \psi^2_1 \rangle
\, , \label{Fsigma} \\
\frac{1}{g^2}
      &=& N_cG \langle \psi_1^2 \rangle 
\,, \label{g} \\
y_1
&=&-y_2 =
-N_cG \cdot 
         \langle 1 + \psi_1 - \psi_0^2 \rangle 
\,, \label{y1} \\
y_3
&=& -y_4=
-N_cG \cdot 
\langle \psi^2_1 \left( 
                   1 + \psi_1 
            \right)^2 
\rangle 
\,, \label{y3} \\ 
y_5
&=&2 y_8=-y_9=
-2N_cG \cdot 
\langle \psi_1^2 \psi_0^2 
\rangle 
\, , \label{y5} \\
y_6
&=&
-y_5-y_7 \,, \label{y6} \\
y_7
&=&
2N_cG \cdot 
\langle \psi_1 \left ( 
                  1 + \psi_1 
          \right) 
          \left( 
                 1 + \psi_1 - \psi_0^2 
          \right)
\rangle 
\,, \label{y7} \\ 
z_4
&=&
-2N_cG \cdot 
\langle \psi_1 \left( 
                 1 + \psi_1 - \psi_0^2 
          \right) 
\rangle 
\,, \label{z4} \\
z_5
&=&
-2N_cG \cdot 
\langle \psi_1^2 \left( 
                   1 + \psi_1 
            \right) 
\rangle 
\,, \label{z5} 
\end{eqnarray}
with $\lambda_1$ being the eigenvalue determined by solving the eigenvalue equation, 
and 
\begin{equation} 
\langle A \rangle \equiv \int dz K^{-1/3}(z) A(z)
\end{equation} 
   for a function $A(z)$. 
In Eq.(\ref{Fsigma}), we used an identity 
\begin{equation}
\int dz K(z) \dot{\psi}_1^2(z) = \lambda_1 \int dz K^{-1/3}(z) \psi_1^2(z) 
\,. 
\end{equation} 
We should note that the normalization of the eigenfunction $\psi_1$ is not solely determined from 
the eigenvalue equation and the boundary condition $\psi_1(\pm \infty)=0$. 
In addition, the values of the 't~Hooft coupling $G$ and the mass scale $M_{KK}$ are not fixed in the model. 
As a result, none of three parameters of the HLS at the leading order, ($F_{\pi}, F_{\sigma}, g$), are fixed 
in the present model: We need three phenomenological inputs to fix their values. 
However, this implies that several physical predictions can be made from only three phenomenological inputs.

It should be also noticed that the Lagrangian (\ref{HLS Lag}) has all the parameters 
consistently with the large $N_c$ counting rule, 
although several ones are absent since the external gauge fields 
are not incorporated in the model: 
As is well known, terms including two or more traces
are suppressed by $1/N_c$ compared with terms of just one trace
in the large $N_c$ limit~\cite{Gas:84}. 
We note that all the terms in Eq.(\ref{HLS Lag}) are 
of $\mathcal{O}(N_c)$ as one can easily see in Eqs.(22)-(31), 
which are constructed by just one trace of the product.

\section{$\rho$-$\pi$-$\pi$ coupling and KSRF~II relation at large $N_c$ limit}

{}As usual in the HLS model~\cite{Bando:1984ej,HY:PRep}, 
from the Lagrangian (\ref{HLS Lag}), we can easily read off the $\rho$ mass square 
and the $\rho$-$\pi$-$\pi$ coupling: 
\begin{eqnarray} 
m_{\rho}^2  & =& a g^2 F_{\pi}^2 
\,, \nonumber \\ 
g_{\rho\pi\pi} &=& \frac{1}{ 2}ag \left( 1 +  \frac{1}{2} g^2 z_4  \right) 
\,, 
\end{eqnarray} 
where phenomenologically important parameter $a$ is defined by
\begin{equation}
a \equiv \frac{F_\sigma^2}{F_\pi^2} \ .
\end{equation} 
It should be noted that these quantities are expressed in terms of the parameters of 
the 5-dimensional gauge theory, by using Eqs.(\ref{Fpi})-(\ref{g}) and (\ref{z4}), as 
\begin{eqnarray} 
m_\rho^2 &=& \lambda_1 M_{KK}^2 
\,, \\
g_{\rho\pi\pi} 
      &=&
       \frac{\pi}{4} \frac{\lambda_1 }{\sqrt{N_c G}} 
       \sqrt{ \frac{ \langle \psi_1(1-\psi_0^2) \rangle^2 }{ \langle \psi_1^2 \rangle}} 
\,. 
\end{eqnarray} 
Since they are independent of the normalization of the eigenfunction $\psi_1$, 
$m_\rho^2$ and $g_{\rho\pi\pi}$ are completely determined, once the values of $G$ and $M_{KK}$ are fixed. 
Moreover, the following ratio related to the KSRF II relation is calculable even independently of these inputs:
\begin{eqnarray}
\frac{m_{\rho}^2}{g_{\rho\pi\pi}^2 F_{\pi}^2} 
        = 
        \frac{4}{a \left( 1 +  \frac{1}{2} g^2 z_4 \right)^2} \nonumber
        &=&
        \frac{4}{\pi} \frac{ \langle \psi_1^2 \rangle }
{ \lambda_1 \langle \psi_1(1 - \psi_0^2) \rangle^2 } \\ 
&\simeq& 3.0
\, , \label{bare KSRFII} 
\end{eqnarray} 
which is roughly 50\% larger than the value of  
the  KSRF II relation, 
\begin{equation} 
\frac{m_{\rho}^2}{g_{\rho\pi\pi}^2 F_{\pi}^2} =2
\,,
\end{equation}
or the experimental value estimated as 
\begin{equation} 
 \frac{m_{\rho}^2}{g_{\rho\pi\pi}^2 F_{\pi}^2} \Bigg|_{\rm exp} = 1.96 
\,, 
\end{equation} 
where use has been made of $F_{\pi}= 92.4\,\,{\rm MeV}$, $m_{\rho} = 775.8 \,\,{\rm MeV} $ 
and $g_{\rho\pi\pi} = 5.99 $. 
Alternatively, when we use $F_{\pi}(0)=86.4$ MeV in the chiral limit~\cite{HY:PRep}, 
\begin{equation} 
 \frac{m_{\rho}^2}{g_{\rho\pi\pi}^2 F_{\pi}^2(0)} \Bigg|_{\rm chi} = 2.24 
\,. 
\end{equation}
The result coincides with 
that in Ref.~\cite{SaSu}. 
This must be so, since different identifications of the $\rho$ meson field, whether the gauge field or 
the CCWZ matter field, cannot lead to different results 
{\it as far as the tree-level amplitude is concerned}~\cite{HY:PRep}.

\section{$1/N_c$-Subleading Corrections}

Now we propose a way to include a part of the $1/N_c$ corrections through meson loops as follows: 
Let us consider the Lagrangian (\ref{HLS Lag}), which has the parameters determined in 
the large $N_c$ limit, as the {\it bare Lagrangian} defined at a scale $\Lambda$: 
$\mathcal{L} = \mathcal{L}(\Lambda)$~\cite{Harada:1999zj,HY:PRep}. 
Then the parameters in the bare Lagrangian are defined as the {\it bare parameters} such as 
$F_{\pi} = F_{\pi}(\Lambda)$, $a=a(\Lambda)$, $g=g(\Lambda)$, and so on. 
The bare theory is matched to the HQCD at the scale $\Lambda$ which we call the matching scale. 
Then, the $1/N_c$ corrections are incorporated into physical quantities in such a way that 
we consider the quantum correction generated from the $\rho$ and $\pi$ loops in HLS ChPT.

 For $m_{\rho}\le \mu \le\Lambda$ the quantum corrections are incorporated through 
the renormalization group equations (RGEs) for 
$F_{\pi}(\mu)$, $a(\mu)$, $g(\mu)$, and $z_4(\mu)$ in the HLS theory {\it including 
the quadratic divergence} in the Wilsonian sense~\cite{Harada:1999zj,HY:PRep}:
\footnote{
           Coefficients of RGEs for all the $\mathcal{O}(p^4)$  terms including $z_4$ 
           are given in Appendix D, Table 20 of Ref. \cite{HY:PRep}.         
          } 
\begin{eqnarray}
\mu \frac{dF_{\pi}^2}{d\mu}
&=&
\frac{N_f}{2(4\pi)^2}
[3a^2g^2F_{\pi}^2 + 2(2-a) \mu^2]
\,, \label{RGE of F pi} \\
\mu \frac{da}{d \mu} 
&=&
-\frac{N_f}{2(4\pi)^2}
(a-1) 
\nonumber \\ 
&& \times \Bigg[ 
              3a(a+1)g^2  
- (3a-1) \frac{\mu^2}{F_{\pi}^2} 
      \Bigg] 
\,,\label{RGE of a} \\
\mu \frac{dg^2}{d \mu} 
&=&
-\frac{N_f}{2(4\pi)^2} \frac{87-a^2}{6}g^4
\,,\label{RGE of g} \\
\mu \frac{dz_4}{d \mu}
&=&
\frac{N_f}{2(4\pi)^2} \frac{2 + 3a - a^2}{6} 
\, . \label{RGE of z4}
\end{eqnarray}
Since
$z_4(\Lambda)$ is related to $(a(\Lambda), g(\Lambda))$ as
\begin{equation}
a(\Lambda) \left(1 + \frac{1}{2} g^2(\Lambda) z_4(\Lambda) \right)^2
 \simeq \frac{4}{3}
\,, \label{a-z4}
\end{equation}
through the HQCD result in Eq.(\ref{bare KSRFII}),
all four parameters in the low-energy region are determined
from just three bare parameters
$F_\pi({\Lambda})$, $g(\Lambda)$ and $a(\Lambda)$
through the above RGEs.
Note that the $\rho$ meson mass $m_\rho$ is determined by the
on-shell condition: 
\begin{equation} 
m_\rho^2 = a(m_\rho) F_{\pi}^2(m_\rho) g^2(m_\rho)
\, .
\end{equation}  
For $0\le \mu \le m_{\rho}$, on the other hand, the couplings other than $F_{\pi}$ do not run, 
while $F_{\pi}$ does by the quantum corrections from the $\pi$ loop alone. 
As a result, the physical decay constant 
$F_{\pi}=F_{\pi}(0)$ is related to $F_{\pi}(m_\rho)$ via the RGEs~\cite{HY:PRep}: 
\begin{eqnarray} 
F_\pi^2(0)
       &=& F_\pi^2(m_\rho) 
            \Bigg[  
                   1 - \frac{N_f}{(4\pi)^2} \frac{m_\rho^2}{F_\pi^2(m_\rho)}  
                   \nonumber \\ 
        &&  \hspace{65pt} \times \left( 1 - \frac{a(m_\rho)}{2} \right)
            \Bigg] 
            \,.  \label{fpichi}
\end{eqnarray}

{}Following Ref.~\cite{HY:PRep},
we take as inputs $N_f=3$, $F_{\pi}(0)=86.4 \pm 9.7$ MeV (value at the chiral limit) and 
$m_{\rho} = 775.8$ MeV~\cite{PDG}, and a particular parameter choice
\footnote{ 
           $a=4/3$ implies the $\rho$ dominance of the $\pi$ $\pi$ scattering~\cite{HY:PRep}.
          } 
\begin{equation}
z_4(\Lambda)=0 \,, \quad {\rm i.e.},  \quad a(\Lambda) \simeq \frac{4}{3} \simeq 1.33 
\, ,
\end{equation} 
among those satisfying  Eq.(\ref{a-z4}), 
 so that $z_4(m_{\rho})$ is solely induced 
by the loop corrections  ($1/N_c$ corrections). 
{}From these, we determine the values of $F_{\pi}(\Lambda)$ and $g(\Lambda)$ 
as done in Ref.~\cite{HY:PRep}. 
We choose the matching scale $\Lambda$ as $\Lambda=1.0,1.1,$ and $1.2$ GeV 
since the effect from the $a_1$ meson is not included.

\begin{table}

\begin{center}

\begin{small} 

\begin{tabular}{ccc}
\hline 
 
\hspace{10pt} $\Lambda\,\,[\rm GeV]$\hspace{10pt} & 
\hspace{10pt} $m_{\rho}^2/(g_{\rho\pi\pi}^2F^2_{\pi})$ \hspace{10pt} &
\hspace{10pt} $g_{\rho\pi\pi}$ \hspace{10pt} 
\\ 
\hline 
 1.0 & $ 1.98 \pm 1.01 $ & $ 6.38 \pm 1.46 $            \\ 
 1.1 & $ 2.01 \pm 1.02 $ & $ 6.34 \pm 1.45 $             \\ 
 1.2 & $ 2.04 \pm 1.04 $ & $ 6.28 \pm 1.44 $             \\ 
\hline

Exp.    & $1.96 \pm 0.00 $  & $ 5.99 \pm 0.03 $ \\ 
Chi.    & $ 2.24 \pm 0.50 $ &  \\ 
\hline 
\end{tabular} 
\vspace{5pt}
\caption{\footnotesize Predicted values for the KSRF II relation and $g_{\rho\pi\pi}$
including the $1/N_c$ corrections with $F_\pi(0)$ and $m_\rho$
used as inputs.
Value of the ratio $m_{\rho}^2/(g_{\rho\pi\pi}^2F^2_{\pi})$
indicated by ``Exp."  is obtained with
the experimental value $F_{\pi}=92.4 \,\, {\rm MeV}$,
while the one by ``Chi." is with the value
$F_{\pi}(0)=86.4 \pm 9.7\,\,{\rm MeV}$ (at the chiral limit).
All errors of the predictions arise from the input value
of $F_{\pi}(0)$. 
         }
\label{rho pi pi} 
\end{small} 

\end{center}

\end{table}

We should carefully define the physical $\rho$-$\pi$-$\pi$
coupling $g_{\rho\pi\pi}$. 
One would naively regard the physical $\rho$-$\pi$-$\pi$ coupling as 
\begin{equation} 
g_{\rho\pi\pi} = \frac{1}{2} a(m_\rho) g(m_\rho) \left[1 + \frac{1}{2} g^2(m_\rho) z_4(m_\rho) \right]
\, ,
\end{equation} 
where $a(m_\rho)=F_\sigma^2(m_\rho)/F_\pi^2(m_\rho)$. 
However, $g_{\rho\pi\pi}$ should be defined for the
 rho meson and the pion both on the mass shell.
While $F_\sigma^2$ and $g$ as well as $z_4$ do not run
for $\mu<m_\rho$, $F_\pi^2$ does run.
Since the on-shell pion decay constant is given by $F_\pi(0)$,
we have to use $F_\pi(0)$ to define the on-shell
$\rho$-$\pi$-$\pi$ coupling constant~\cite{HY:PRep}.
The resultant expression is given by
\begin{equation}
g_{\rho\pi\pi} = \frac{1}{2} a(0) g(m_\rho)
         \left(  1 + \frac{1}{2} g^2(m_\rho) z_4(m_\rho)  \right)
\, , \label{grhopipi}
\end{equation}
where  $a(0) \equiv F_{\sigma}^2(m_{\rho})/F_{\pi}^2(0)$
is related to $a(m_\rho)$ through Eq.(\ref{fpichi}) as:
 \begin{eqnarray} 
\frac{1}{a(0)} 
       &=& \frac{1}{a(m_\rho)} 
            \Bigg[  
                   1 - \frac{3}{(4\pi)^2} \frac{m_\rho^2}{F_\pi^2(m_\rho)} 
\nonumber \\ 
                   && \hspace{65pt} \times  
                    \left( 1 - \frac{a(m_\rho)}{2} \right)
            \Bigg] 
            \, .  \label{azero}
\end{eqnarray}  

   By using the above $g_{\rho\pi\pi}$, the physical quantity  related to the KSRF~II relation is given by 
\begin{eqnarray} 
\frac{m_{\rho}^2}{g_{\rho\pi\pi}^2 F_{\pi}^2(0)} 
        &=& 
         \frac{4}{ a(0) \left(1 + \frac{1}{2} g^2(m_{\rho}) z_4(m_{\rho}) \right)^2} \nonumber \\
         &\simeq& 2.0 \, ,
 \label{on-shell g rho pi pi }
\end{eqnarray}  
in good agreement with the experiment, where we have computed 
\begin{equation} 
a(0) \simeq 2.0  \quad , \quad  
\frac{1}{2} g^2(m_\rho) z_4(m_\rho) \simeq -8.0 \times 10^{-3}
\end{equation} 
 through RGE analysis for $\Lambda=1.1 {\rm GeV}$, which are compared with
 the bare values $a(\Lambda)\simeq 4/3$ and $\frac{1}{2} g^2(\Lambda) z_4(\Lambda)=0$. Eq. (\ref{on-shell g rho pi pi }) is our main result, which is compared with the holographic result Eq.(\ref{bare KSRFII}).
 
  We note that those corrections are of $\mathcal{O}(1/N_c)$. Actually,
we may set $a(m_\rho)\simeq a(\Lambda)$ in Eq.(\ref{azero}), since 
 $a(\mu)$ does hardly run for $m_\rho < \mu  < \Lambda$ due to the fact that 
the bare value $a(\Lambda) \simeq 1.33$ is close to the fixed point value $a=1$ of the RGE (\ref{RGE of a}) 
(See also Fig.~17 of Ref.~\cite{HY:PRep}). Then  
 \begin{eqnarray} 
\frac{1}{a(0)} 
       &\simeq& \frac{1}{a(\Lambda)} 
            \Bigg[  
                   1 - \frac{3}{(4\pi)^2} \frac{m_\rho^2}{F_\pi^2(m_\rho)}  
                   \nonumber \\ 
                   && \hspace{65pt} \times
                    \left( 1 - \frac{a(\Lambda)}{2} \right)
            \Bigg] 
              \label{azero rev}
\end{eqnarray}  
whose  second term in the bracket with $m_\rho^2/F_\pi^2$ ($\sim 1/N_c$) is nothing but 
  the $\mathcal{O}(1/N_c)$ correction essentially coming from the pion loop contributions 
  for $0 < \mu < m_\rho$.

In Table~\ref{rho pi pi}, we show the predicted values of $m_{\rho}^2/(g_{\rho\pi\pi}^2F^2_{\pi}(0))$ and of 
$g_{\rho\pi\pi}$ for $\Lambda=1.0, 1.1,1.2$ GeV in good agreement with the experiment 
within the errors coming from the input value $F_{\pi}(0)$ evaluated at the chiral limit~\cite{HY:PRep}. 
The result is fairly insensitive to the choice of the matching scale $\Lambda$. 
This implies that $1/N_c$ corrections actually improve the HQCD prediction, Eq.(\ref{bare KSRFII}), 
$ m_{\rho}^2/(g_{\rho\pi\pi}^2 F_{\pi}^2)|_\Lambda \simeq 3.0$, into the realistic value $\simeq 2.0$. 
It should be emphasized that the $1/N_c$ corrections make the value always closer to the experimental value 
for a wide range of the value of the parameter $a(\Lambda)$ not restricted to the present one $a(\Lambda)\simeq 4/3$. 
\\

By introducing external field, SS~\cite{SaSu} 
obtained  
``vector meson dominance'' for the pion electromagnetic form factor, 
though not the celebrated ``$\rho$ dominance'' due to significant contributions 
from  higher resonances, particularly the $\rho^\prime$. 
The above peculiarity is closely related to its prediction of $g_\rho$, 
the $\rho$-$\gamma$ mixing strength, or the pion form factor just on the $\rho$ pole in the {\it time-like region}, 
namely a wrong KSRF I relation, $g_\rho/(g_{\rho\pi\pi} F_\pi^2) \simeq 4$~\cite{SaSu}, 
which is a factor 2 larger than the correct one. 
These problems will be dealt with in the forthcoming paper~\cite{forthcoming}.

\section*{Acknowledgements}

We would like to thank Sekhar Chivukula, Shigeki Sugimoto and Masaharu Tanabashi for 
useful comments and discussions. 
This work was supported in part by The Mitsubishi Foundation and 
JSPS Grant-in-Aid for Scientific Research (B) 18340059, and by the 21st Century COE Program of 
Nagoya University provided by JSPS (15COEG01). 
It was also supported in part by the Daiko Foundation \#9099 (M.H.) and JSPS Grant-in-Aid for 
Scientific Research (C)(2) 16540241 (M.H.).

\end{document}